\documentstyle[prl,aps,twocolumn,epsf,floats]{revtex}
\def\geqap{\,\raise 2pt \hbox{$>\kern-11pt \lower 5pt \hbox{$\sim$}$}\,}
\def\leqap{\,\raise 2pt \hbox{$<\kern-10pt \lower 5pt \hbox{$\sim$}$}\,}
\begin{document}
\draft
\twocolumn[\hsize\textwidth\columnwidth\hsize\csname @twocolumnfalse\endcsname
\title{Orbital Ordering and Resonant X-ray Scattering in Layered Manganites}
\author{Sumio Ishihara and Sadamichi Maekawa}
\address{Institute for Materials Research, Tohoku University,  Sendai,
980-8577, Japan}

\date{\today}
\maketitle
\begin{abstract} 
In layered manganites with orbital and charge orderings, 
the degeneracy of the Mn $4p$ orbitals as well as 
the $3d$ ones is lifted by the effects of the $4p$ bands and 
the local Coulomb interactions. 
We formulate the atomic scattering factor for the resonant 
x-ray scattering in the memory function method 
by taking into account these 
effects on an equal footing. 
It is shown that the polarization dependences of the scattering intensities 
at the orbital and charge superlattice reflections observed in 
LaSr$_{2}$Mn$_2$O$_7$ are 
caused by the local and itinerant characters of $4p$ 
electrons, respectively. 
We examine the type of the orbital ordered state. 

\end{abstract}
\pacs{PACS numbers: 75.30.Vn, 71.10.-w, 78.70.Ck} 
]

\narrowtext
Orbital degree of freedom is one of the 
key factors to uncover the dramatic and fruitful phenomena 
in colossal magnetoresistive (CMR) manganites. 
However, since the experimental technique to  
detect the orbital structure has been limited for a long time \cite{ito}, 
it was considered to be a hidden degree of freedom. 
The resonant x-ray scattering (RXS) has recently shed light on this issue. 
Murakami $\it et~al.$ have applied the scattering method 
to La$_{0.5}$Sr$_{1.5}$MnO$_4$ 
and observed the superlattice reflections 
near the Mn$^{3+}$ $K$-edge caused by the orbital 
ordering \cite{murakami1}. 
Immediately after the observation, this method was 
extensively studied in both the experimental and theoretical sides 
\cite{murakami2,endoh,nakamura,hill1,ishihara3,elfimov} 
and has been recognized to be a powerful tool to detect the orbital states 
through the application to several manganites. 
\par
Near the Mn$^{3+}$ $K$-edge, RXS  
is given by the electric dipole transition 
between Mn $1s$ and $4p$ orbitals. 
The anomalous part of the atomic scattering factor (ASF) 
is a tensor with respect to the polarization of x-ray 
and the anisotropy of the tensor elements 
brings about the scattering at the orbital superlattice reflections. 
It was shown that the Coulomb interactions between Mn $3d$ and Mn $4p$ electrons  
and between Mn $3d$ and O $2p$ ones are the convincing candidates to 
induce the anisotropy 
in the orbital ordered states \cite{ishihara3}. 
\par
In this letter, 
we focus on RXS in manganites with layered structure. 
The layered manganites are recognized to be materials appropriate 
for studying orbital degree of freedom as well as CMR \cite{moritomo}. 
In double layered manganites La$_{2-2x}$Sr$_{1+2x}$Mn$_2$O$_7$, 
the orbital states are controlled by changing $x$ and  
applying pressure and magnetic field \cite{kimura,argyriou,ishihara4}. 
The orbital states of $3d$ electrons 
as well as $4p$ ones depend on the deformation of the MnO$_6$ octahedron 
and the layered crystal structure. 
However, the octahedron is almost isotropic in the 
double layered manganites with $x=0.5$ \cite{kubota} 
where the orbital and charge orderings were recently 
observed by RXS \cite{wakabayashi}. 
Thus, the energy levels of the three $4p$ orbitals  
are nearly degenerate in the local sense. 
On the other hand, the layered structure 
provides the quasi-two dimensional character of the $4p$ band 
and lifts the degeneracy. 
We formulate ASF in the memory function method 
by taking into account 
the effects of the $4p$ band and 
the local Coulomb interactions on an equal footing.  
The anisotropy of ASF 
reflected from the local and itinerant characters of the 
$4p$ electrons explains 
the experimental results in LaSr$_{2}$Mn$_2$O$_7$. 
Being based on the calculated results, 
types of the orbital ordered state are discussed. 
\par
Let us first formulate ASF. 
On resonance, the relevant term in the diagonal part of ASF is 
given by \cite{blume,ishihara3} 
\begin{eqnarray}
\Delta f_{\alpha}&=&
{m \over e^2 } \sum_{l} 
{  \langle  0 | j_{i \alpha}  | l \rangle 
  \langle l | j_{i \alpha}    | 0 \rangle 
  \over 
  \varepsilon _0-\varepsilon_l+\omega-i\Gamma } 
  \ , 
\label{eq:dltf}    
\end{eqnarray}
where $|0 \rangle$ ($|l \rangle$) denotes 
the initial (intermediate) electronic state with energy 
$\varepsilon_0$ ($\varepsilon_l$).  
$\omega$ is the energy of x-ray with 
polarization $\alpha$. 
$j_{i \alpha}(={e \over m} A
\sum_\sigma P^\dagger_{i \alpha \sigma} s_{i \sigma}+H.c.)$ 
describes the $1s \rightarrow 4p$ transition where 
$P^\dagger_{i \alpha \sigma}$ and $s_{i \sigma}$ 
are the creation operator of Mn $4p$ electron at site $i$ with 
spin $\sigma$ and orbital $\alpha(=x,y,z)$ and the annihilation 
one of Mn $1s$ electron, respectively. 
$\Gamma$ denotes the damping of a core hole. 
Because the radius of the Mn $1s$ orbital is much smaller 
than the lattice constant, 
the $1s$ electron is excited to the $4p$ orbital at the same site. 
The relevant part of $\Delta f_{\alpha}$ 
is rewritten by using the Green's function for the operator 
$J_{i \alpha \sigma}=P^\dagger_{i \alpha \sigma } s_{i \sigma}$ 
as follows,  
\begin{equation}
{\rm Im}\Delta f_{\alpha}=-{e^2 |A|^2 \over m} 
\sum_\sigma {\rm Im}G_{\alpha \sigma}(z) \Big |_{z=\omega+i\Gamma} \ , 
\label{eq:green}
\end{equation}
%
%
where $G_{\alpha \sigma}(z)$ is the Fourier transform (FT) of 
the retarded Green's function 
$G_{\alpha \sigma}(t)={i \over 2 \pi} \int e^{-i \omega t} 
G_{\alpha \sigma}(\omega) d\omega$
defined by 
$
G_{\alpha \sigma}(t)=\theta(t)\langle [ J_{i \alpha \sigma}^\dagger(t) , J_{i \alpha \sigma}(0) ] \rangle \ . 
$
%
\par
We consider a MnO$_6$ cluster where 
x-ray is absorbed.  The electrons in the cluster 
couple to the $4p$ bands. 
The Hamiltonian is given by 
%
$
H=H_{0}+H_{4p} \ , 
$
with  
\begin{eqnarray}
H_{0}&=&\varepsilon_d \sum_{\gamma \sigma} d^\dagger_{i \gamma \sigma} d_{i \gamma \sigma}
   +\varepsilon_p \sum_{\gamma \sigma} p^\dagger_{i \gamma \sigma} p_{i \gamma \sigma}
   +\varepsilon_s \sum_{\sigma}        s^\dagger_{i \sigma}        s_{i \sigma} 
   \nonumber \\
   &+&t_{pd} \sum_{\gamma \sigma} (d^\dagger_{i \gamma \sigma} p_{i \gamma \sigma}
                                  +p^\dagger_{i \gamma \sigma} d_{i \gamma \sigma} )
                                  \nonumber \\
   &+&U\sum_\gamma n_{i \gamma \uparrow}^d n_{i \gamma \downarrow}^d
   +U'\sum_{\sigma \sigma'} n_{i a \sigma}^d n_{i b \sigma'}^d
   \nonumber \\
   &+&I\sum_{\sigma \sigma'} d^\dagger_{i a \sigma} d^\dagger_{i b \sigma'} 
                             d_{i a \sigma'} d_{i b \sigma} 
   -J_H \vec S_i \cdot \vec S_{t i}
   \nonumber \\
   &+&n_{h} \sum_{\gamma \sigma}( V_{sd} n^d_{i \gamma \sigma}
                                + V_{sp}  n^P_{i \alpha \sigma} )
    +\sum_{\alpha \gamma \sigma \sigma'} V_{\gamma \alpha}
     n_{i \gamma \sigma}^d n_{i \alpha \sigma' }^P \ ,
\label{eq:hcluster}
\end{eqnarray}
and 
\begin{eqnarray}
H_{4p}=\varepsilon_P \sum_{j \alpha \sigma} P^\dagger_{j \alpha \sigma}
                                              P_{j \alpha \sigma} 
+\sum_{j \delta \alpha \beta \sigma} 
t_\alpha^\beta P^\dagger_{j \alpha \sigma} P_{j+\delta_\beta \alpha \sigma} \ ,  
\label{eq:h4p}
\end{eqnarray}
where $H_0$ and $H_{4p}$ describe 
the electronic states in the MnO$_6$ cluster 
and in the $4p$ band, respectively \cite{hamil}. 
$i$ indicates the site where x-ray is absorbed. 
$d_{i \gamma \sigma}$  is 
the annihilation operator 
of Mn $3d$  electron with spin $\sigma$ and orbital 
$\gamma(=3z^2-r^2, x^2-y^2)$ and 
$p_{i \gamma \sigma}$ is that of O $2p$ electron 
with $\sigma$ and $\gamma$ represented by 
combining six O $2p$ orbitals 
in a MnO$_6$ octahedron \cite{ishihara3}. 
The number operator 
of Mn $3d$ $(4p)$ electrons and 
that of core holes are denoted by 
$n_{i \gamma \sigma}^d$ $(n_{i \alpha \sigma}^P)$ and $n_{i h}$, respectively. 
The last three terms in Eq.~(\ref{eq:hcluster})
describe the 
the core hole potentials and the Coulomb interaction 
between $3d$ and $4p$ electrons, respectively. 
The explicit form of $V_{\gamma \alpha}$ is given by  
$V_{\gamma \alpha}=F_0+4F_2\cos(\theta_\gamma+m_\alpha {2 \pi \over 3})$ 
with the angle in the orbital space 
$\theta_\gamma$ defined by 
$|\theta_\gamma \rangle=\cos(\theta_\gamma/2)|d_{3z^2-r^2} \rangle+\sin(\theta_\gamma/2)| d_{x^2-y^2} \rangle$
and $(m_x,m_y,m_z)=(1,2,3)$ \cite{ishihara3}. 
This interaction directly connects the orbital states of $3d$ electrons with 
that of $4p$ ones. 
The crystal with layered structure is 
described by the tetragonal lattice in $H_{4p}$. 
$t_\alpha^\beta$ is the hopping integral between 
the site $j$ with orbital $\alpha$ 
and its nearest neighboring site $j+\delta_\beta$ with $\alpha$, 
where $\delta_\beta$ indicates a bond in the $\beta$ direction.  
The following conditions are satisfied:  
$t_x^x=t_y^y \equiv t_\sigma^\parallel$, $t_z^x=t_z^y=t_x^y=t_y^x \equiv t_\pi^\parallel$, 
$t_z^z \equiv t_\sigma^\perp$ and $t_x^z=t_y^z \equiv t_\pi^\perp$.   
The effect of the lattice distortion \cite{elfimov} is neglected, 
since it is experimentally confirmed to be  
small in the layered manganites 
around $x=0.5$ \cite{kubota}. 
Thus, $\varepsilon_{d(p)}$ and $\varepsilon_P$ are chosen to be 
independent of $\gamma$ and $\alpha$, respectively. 
The inter-site Coulomb interaction 
between Mn $4p$ and O $2p$ electrons is not included, 
because the interaction brings about 
the anisotropy of ASF in the similar way as $V_{\gamma \alpha}$ \cite{ishihara3}. 
\par
The memory function method (the composite operator method) 
in the Green's function formalism is adopted in the calculation of ASF. 
This method 
is suitable to describe 
the excitation in the highly correlated systems \cite{forster,matsumoto} 
and treats the local Coulomb interactions and 
the $4p$ band on an equal footing. 
We introduce the relaxation function defined by 
%
$
C_{\alpha \sigma}(t)$
$
=\theta(t) \beta^{-1} \int^\beta_0 d\lambda
\langle J^\dagger_{i \alpha \sigma}(t) J_{i \alpha \sigma}(i\lambda) \rangle$
$ 
\equiv \theta(t) \langle J_{i \alpha \sigma}^\dagger(t) J_{i \alpha \sigma}(0) \rangle_{\lambda} 
$  
with $\beta=1/T$.  
We have 
${\rm Im} C_{\alpha \sigma}(\omega)={\rm Im}G_{\alpha \sigma}(\omega)/(\beta \omega)$, 
$C_{\alpha \sigma}(\omega)$ being FT of 
$C_{\alpha \sigma}(t)$.  
At the end of the calculation, we take $T=0$. 
The final form of 
$C_{\alpha \sigma}(\omega)(\equiv \delta M_C^{(0)}(\omega))$ is 
given by the continued fraction \cite{matsumoto}: 
\begin{equation}
\delta M^{(n-1)}_C (\omega)={ I^{(n)} \over 
\omega-\bigl( M_0^{(n)}+\delta M^{(n)}(\omega) \bigr) I^{(n)-1}  } \ , 
\end{equation}
with 
%
$
\delta M^{(n)}(\omega)=\delta M^{(n)}_B(\omega)+\delta M^{(n)}_C(\omega) 
$ 
%
%
for $n > 1$. 
$\delta M_C^{(n-1)}(\omega)$ is given by FT of 
$\delta M_C^{(n-1)}(t)=\theta(t) \langle \psi_n^{\dagger}(t) \psi_n(0) \rangle_{\lambda}$.  
$\delta M^{(3)}(\omega)$ is taken to be zero 
and the continued fraction is truncated at $n=3$. 
Here, $\psi_n$ is the composite operator represented
by an operator product \cite{composite}. 
$\psi_1=J_{i \alpha \sigma}$ 
and $\psi_{n+1}$ appears in the 
equation of motion of $\psi_n$.  
$I^{(n)}(=\langle \psi_n^{\dagger} \psi_n \rangle_{\lambda})$ and  
$M_0^{(n)}(=\langle (i\partial_t \psi_n^{\dagger}) \psi_n \rangle_{\lambda})$ 
correspond to the normalization factor and the static part of 
the self-energy, respectively. 
$\delta M_B^{(n)}(\omega)$ 
is the relaxation function  
including operators at site $j \ne i$, 
such as $P_{j \alpha \sigma}^\dagger s_{i \sigma}$.  
The detailed formulation will be presented elsewhere. 
Advantages of this method in the present issue 
are the following: 
1) 
The many body excitations originating from the local Coulomb interactions 
are treated by the composite operators.
For example, 
$P^\dagger_{i \alpha \sigma} s_{i \sigma} 
d_{i \gamma \sigma'}^\dagger p_{i \gamma \sigma'}$  
describes the dipole transition associated with the 
charge transfer between Mn $3d$ and O $2p$ orbitals.  
By treating $\psi_n$ as a single quantum variable, 
the many body excitations and 
the interactions between them, i.e. so-called 
the configuration interactions, 
are taken into account in $\delta M^{(n)}_C(\omega)$ and $M_0^{(n)}$.  
This kind of the excitation is not treated by the independent 
single-particle scheme with an averaged potential.
2) 
The band effects of the $4p$ electrons are included in 
$\delta M^{(n)}_B(\omega)$. 
By adopting the loop approximation in the diagram 
technique, where the $4p$ state is treated as 
the single-particle state described by $H_{4p}$, 
the itinerant character of the excitation 
is introduced. 
This kind of excitation plays a crucial role on ASF in the present 
case where the layered structure lifts the orbital 
degeneracy and is not described by the calculation 
in a small cluster. 
\par
%
\begin{figure}
\epsfxsize=6.5cm
\centerline{\epsffile{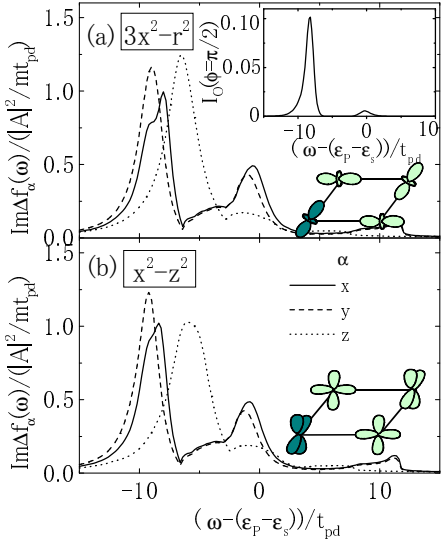}}
%
%
\caption{
The imaginary part of ASF. 
(a) The $d_{3x^2-r^2}$ orbital is occupied in the $[d_{3x^2-r^2}/d_{3y^2-r^2}]$-type 
orbital ordered state. The straight, dashed and dotted lines show 
ASF for $\alpha=x$, $y$ and $z$, respectively. 
(b) The $d_{x^2-z^2}$ orbital is occupied in the $[d_{z^2-x^2}/d_{y^2-z^2}]$-type 
orbital ordered state. The insets show the scattering intensity at the orbital 
superlattice reflection and schematic pictures of the orbital ordered states. 
}
\label{fig1}
\end{figure}
The calculated ASF is presented in 
Fig.~1(a), where the $d_{3x^2-y^2}$ orbital is occupied.  
The $[d_{3x^2-r^2}/d_{3y^2-r^2}]$-type 
orbital ordered states, where the two orbitals
are alternately aligned in the $xy$ plane, is assumed. 
The anisotropy of the $4p$ band is characterized by  
$t_\sigma^\parallel/t_\sigma^\perp=4$ and  
$t_\sigma^\parallel/t_\pi^\parallel=t_\sigma^\perp/t_\pi^\perp=4$ \cite{parameter}.
A continuous spectrum in Im$\Delta f_{x(y)}$ 
spreads over a wide region of $\omega$. 
The weight near the $K$-edge is dominated by 
Im$\Delta f_{x(y)}$, because the $4p_{x(y)}$ band 
is wider than the $4p_z$ band; 
the quasi-two dimensional band  of the $4p$ electrons 
causes the anisotropy between $\Delta f_{x(y)}$ and $\Delta f_z$. 
However, 
it is worth to note that 
Im$\Delta f_{\alpha}$ is not the density of states itself. 
There are several peak structures 
in Im$\Delta f_{\alpha}$ caused by  
the local excitations in the MnO$_6$ cluster.
Near the edge, the clear anisotropy between Im$\Delta f_x$ and Im$\Delta f_y$ appears 
and Im$\Delta f_y$ governs the intensity. 
This anisotropy is caused by the Coulomb interaction  
between $3d$ and $4p$ electrons.  
The core hole potential makes the anisotropy remarkable, 
since the potential reduces the energy of the dipole transition and 
enhances the local character 
of the $4p$ electrons. 
The scattering intensity 
at the orbital superlattice reflection defined by   
$I_O(\phi=\pi/2)=|\Delta f_x-\Delta f_y|^2/(2|A|^2/mt_{pd})^2$ 
\cite{ishihara3,pisigma}
is shown in the inset of Fig.~1(a).
Here, the azimuthal angle ($\phi$) is the rotating one of the sample 
around the scattering vector.
For $\phi=0$ ($\pi/2$), 
the electric vector of x-ray is perpendicular (parallel) to the $xy$ plane.  
A sharp peak near the edge together with 
a small intensity above the edge appear in $I_O(\phi)$. 
Both structures are observed  
in LaSr$_{2}$Mn$_2$O$_7$ \cite{wakabayashi}. 
In Fig.~1(b), we show 
ASF at the site 
where the $d_{x^2-z^2}$ orbital is occupied in the 
$[d_{x^2-z^2}/d_{y^2-z^2}]$-type orbital ordered state.
The anisotropy between Im$\Delta f_{x(y)}$ and 
Im$\Delta f_z$ becomes more remarkable near the edge 
due to the Coulomb interaction. 
\par
The anisotropies of ASF directly reflect on the polarization dependence  
of the scattering intensity. 
Let us consider the charge and orbital 
ordered states realized in the layered 
manganite at $x=0.5$ \cite{murakami1,wakabayashi}.
The scattering intensities at the 
orbital and charge reflection points in the 
$[\theta/ -\theta]$-type orbital ordered state  
are given by 
$I_O(\phi)=|(\Delta f_{x}-\Delta f_{y})\sin\phi|^2/(2|A|^2/mt_{pd})^2$ 
and 
$I_C(\phi)=|(\Delta f_{x}+\Delta f_{y})\sin^2\phi
+2\Delta f_{z} \cos^2 \phi|^2/(2|A|^2/mt_{pd})^2$, 
respectively \cite{ishihara3,pisigma}. 
The polarization dependences of 
$I_O(\phi)$ and $I_C(\phi)$ are 
attributed to the anisotropies between 
$\Delta f_x$ and $\Delta f_y$ due to the local Coulomb interactions 
and between $\Delta f_x+\Delta f_y$ and 
$\Delta f_z$ due to the effects of the $4p$ band, 
respectively; the itinerant and local characters of the excited $4p$ 
electrons reflect on $I_C(\phi)$ and $I_O(\phi)$, respectively. 
The theoretical results of $I_O(\phi)$ and $I_C(\phi)$ near the edge are plotted 
together with the experimental data in LaSr$_{2}$Mn$_2$O$_7$ \cite{wakabayashi}
in Fig.~2. 
We find good agreement with theory and experiment. 
Thus, we conclude that ASF in LaSr$_{2}$Mn$_2$O$_7$ 
near the edge is dominated by $\Delta f_{x(y)}$  
and the degeneracy between $\Delta f_x$ and $\Delta f_y$ 
is lifted by the Coulomb interactions 
in the $[\theta /-\theta]$-type orbital ordered state. 
On the other hand, the $\theta$-dependence of ASF 
is not so remarkable as to determine the value of $\theta$. 
However, from the following facts, 
we deduce that 
the $[d_{3x^2-r^2}/d_{3y^2-r^2}]$-type $(\theta=2\pi/3)$
ordered state is more favorable in LaSr$_{2}$Mn$_2$O$_7$
rather than the $[d_{x^2-z^2}/d_{y^2-z^2}]$-type one; 
In the layered structure, 
the $[d_{3x^2-r^2}/d_{3y^2-r^2}]$-type ordered state 
gains more kinetic energy than the $[d_{x^2-z^2}/d_{y^2-z^2}]$-type one, 
because the $3d$ band in the former state  
is wider than that in the latter,  
as discussed in the case of the $4p$ band.  
This is consistent with the zigzag-type 
ferromagnetic spin alignment observed in the 
$xy$ plane \cite{kubota}. 
In the $[d_{3x^2-r^2}/d_{3y^2-r^2}]$-type 
ordered state, the hopping integral in the $z$ direction 
is small , where the same kinds of charge and 
orbital are stacked \cite{wakabayashi}. 
As a result, the double exchange interaction 
in this direction is suppressed and the superexchange 
interaction between $t_{2g}$ spins causes the 
antiferromagnetic spin alignment 
as observed experimentally \cite{kubota}. 
\par
%
%
%
\begin{figure}
\epsfxsize=6.5cm
\centerline{\epsffile{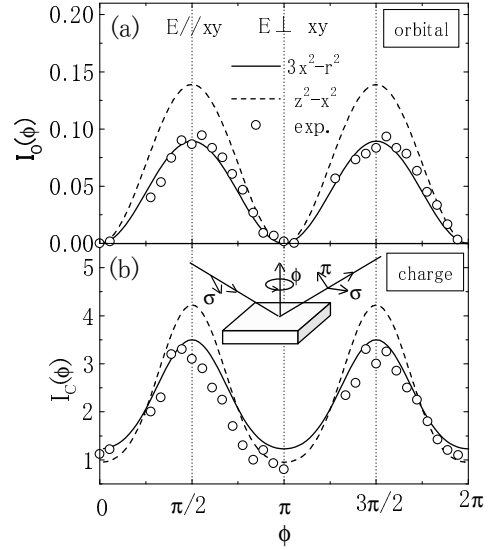}}
%
%
\caption{
The polarization dependence of 
the scattering intensities at the orbital superlattice reflection (a) and 
at the charge superlattice one (b) near the edge 
($(\omega-(\varepsilon_P-\varepsilon_s))/t_{pd}=-9.6$). 
The straight and broken lines show the intensities in the  
$[d_{3x^2-r^2}/d_{3y^2-r^2}]$- and 
$[d_{x^2-z^2}/d_{y^2-z^2}]$-type orbital ordered states, respectively. 
The open circles show the experimental data obtained in 
LaSr$_2$Mn$_2$O$_7$ [14]. 
Absolute values of the experimental data are arbitrary. 
The inset shows the schematic picture of the experimental arrangement.
}
\label{fig2}
\end{figure}
The anisotropy of ASF in the layered structure 
is quite different from that in the cubic structure 
where the $4p$ band dose not contribute to the anisotropy. 
ASF calculated on the condition 
$t_\sigma^\parallel/t_\sigma^\perp
=t_\pi^\parallel/t_\pi^\perp=1$ is shown in Fig.~3 
where the $d_{3x^2-r^2}$ orbital is occupied. 
$\Delta f_y$ and $\Delta f_z$ are almost degenerate and 
the anisotropy between $\Delta f_{y(z)}$ and $\Delta f_x$ 
is provided by the local Coulomb interaction, 
as shown in the previous results 
obtained in the MnO$_6$ cluster \cite{ishihara3}. 
The energy and polarization dependences of $I_O$ 
are similar to those in the inset of 
Fig.~1(a) and Fig.~2(a), respectively. 
$I_C(\phi)$ depends on types of the orbital ordered state 
(the inset of Fig.~3) \cite{cubic,ishihara3}. 
\par
%
%
\begin{figure}
\epsfxsize=6.5cm
\centerline{\epsffile{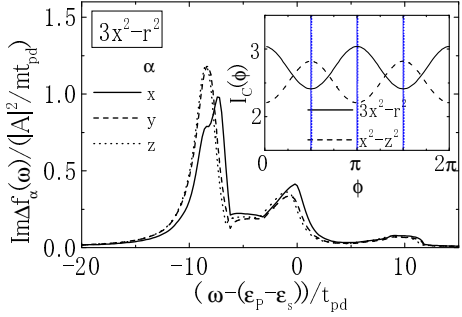}}
%
%
\caption{
The imaginary part of ASF.  
The isotropic $4p$ band is assumed. 
The $d_{3z^2-r^2}$ orbital is occupied in the $[d_{3x^2-r^2}/d_{3y^2-r^2}]$-type 
orbital ordered states. The straight, dashed and dotted lines show 
ASF for $\alpha=x$, $y$ and $z$, respectively.
Im$\Delta f_y$ and Im$\Delta f_z$ are almost degenerate. 
The inset shows 
the polarization dependence of the scattering intensity 
at the charge superlattice reflection $I_C(\phi)$. 
The straight and broken lines show  $I_C(\phi)$ for the 
$[d_{3x^2-r^2}/d_{3y^2-r^2}]$- and 
$[d_{x^2-z^2}/d_{y^2-z^2}]$-type orbital ordered states, respectively. 
}
\label{fig3}
\end{figure}
The anisotropy of ASF in the layered manganites 
is highly in contrast to that in the layered nickelates and cuprates. 
In the latter compounds, the $K$-edge is dominated by the $z$ component 
and the $x(y)$ component is located above $2 \sim 10$ eV from the 
edge \cite{sahiner,hill2}. 
Since the octahedron is elongated along the $z$ axis 
(more than 20$\%$ in cuprates and 15$\%$ in nickerates), 
the large hybridization between $4p_{x(y)}$ and O$2p$ orbitals  
pushes the $4p_{x(y)}$ band to the higher energy region. 
On the contrary,  
in the layered manganites around $x=0.5$, 
the elongation is less than 1$\%$ \cite{kubota} and the orbital degree of freedom 
for the $4p$ electrons survives in the local sense as well as that for the $3d$ ones. 
Thus, contribution from the quasi-two dimensional band  
is relevant near the edge. 
\par
To conclude, we have studied RXS 
in layered manganites by taking into account 
the effects of the $4p$ bands and 
the local Coulomb interactions. 
The anisotropies 
between $\Delta f_{x(y)}$ and $\Delta f_z$ 
and between $\Delta f_x$ and $\Delta f_y$ 
are reflected from the 
local and itinerant characters of the $4p$ electrons, respectively. 
The calculated results well reproduce the experimental 
ones in LaSr$_{2}$Mn$_2$O$_7$. 
\par
\medskip
Authors thank Y.~Murakami and Y.~Wakabayashi 
for providing the experimental data prior to publication,  
T.~Arima and Y.~Endoh for their valuable discussions.  
The work was supported by CREST and NEDO. 
Part of the numerical calculation was performed in the HITACS-3800/380 
superconputing facilities in IMR, Tohoku Univ. 

\vfill
\eject

\begin{references}
\bibitem{hamil}
$\vec S_i$ ($\vec S_{t i}$) is the spin operator for the $e_g$ ($t_{2g}$) electrons. 
$U$ ($U'$), $I$ and $J_H$ are 
the intra (inter)-orbital Coulomb interaction, the exchange interaction and 
the Hund coupling between $e_g$ and $t_{2g}$ spins, respectively.
\bibitem{composite}
The three operators:  
$(\psi_1, \psi_2,\psi_3)=$
$(J_{i \alpha \sigma}$, 
$P_{i \alpha \sigma}^\dagger s_{i \sigma} \delta n_{i \gamma }^d$, 
$P_{i \alpha \sigma}^\dagger s_{i \sigma} \delta n^{dp}_{i \gamma } )$
are adopted as the composite operators, where 
$\delta A=A-\langle A \rangle$ and 
$n^{dp}_{i \gamma}=\sum_{\sigma }d_{i \gamma \sigma}^\dagger p_{i \gamma \sigma}$.  
$\delta M_B^{(n)}(\omega)$ is the relaxation 
function of the operators: 
$(P_{j \alpha \sigma}^\dagger s_{i \sigma}$, 
$P_{j \alpha \sigma}^\dagger s_{i \sigma} \delta n_{i \gamma }^d$, 
$P_{j \alpha \sigma}^\dagger s_{i \sigma} \delta n_{i \gamma}^{dp} )$ for $n=(1,2,3)$ 
where $j$ indicates the nearest neighboring site of $i$.
\bibitem{parameter}
The other parameter values are chosen to be $\varepsilon_p-\varepsilon_d=1.5$, 
$U=5$, $U'=4$, $J_H=I=1$, $V_{sd}=-5.0$, $V_{sp}=-3.5$, 
$F_0=3.5$, $F_2=0.25$, $t_\sigma^\parallel=3$ 
and $\Gamma=1.5$ in units of $t_{pd}$ 
with $t_{pd}$ being about 1eV.
\bibitem{pisigma}
The polarization of the scattered x-ray 
is chosen to be the $\pi$ ($\sigma$) 
component for $I_O(\phi)$ ($I_C(\phi)$). 
Contribution from the Mn$^{4+}$ ion to $I_C(\phi)$
is neglected, since it is small 
near the Mn$^{3+}$ $K$-edge.
\bibitem{cubic}
We note that manganites 
with cubic structure and 
the charge ordering at $x \sim 0.5$  
have a two dimensional character which provides  
the anisotropy between $\Delta f_{x(y)}$ and $\Delta f_{z}$,  
as shown in the layered case.
\end{references}
\end{document}